\documentclass[12pt]{article}

\usepackage{paper2e}
\usepackage{mydefs2e}
\usepackage{xspace}
\usepackage{graphicx} 

\newcommand{\nc}{N_{\it c}}
\newcommand{\nf}{N_{\it f}}

\begin{document}

\begin{titlepage}

\title{Strong Conformal Dynamics at the LHC\\\medskip
and on the Lattice}

\author{Markus A. Luty}

\address{Physics Department, University of California Davis\\
Davis, California 95616}

\begin{abstract}
Conformal technicolor is a paradigm for new physics at the LHC
that may solve the problems of strong electroweak
symmetry breaking for quark masses and precision
electroweak data.
We give explicit examples of 
conformal technicolor theories
based on a QCD-like sector.
We suggest a practical
method to test the conformal dynamics
of these theories on the lattice.
\end{abstract}

\end{titlepage}

\section{Conformal Technicolor}
Electroweak symmetry breaking by strong dynamics at the
TeV scale has great theoretical appeal, but has well-known
difficulties with fermion masses and precision electroweak data.
Conformal technicolor \cite{CTC} is a framework closely
related to ``walking'' technicolor \cite{walking} in which
there are good reasons to think that these problems can be solved.
Whether they are in fact solved depends on the dynamics
of strongly-coupled small-$N$ conformal field theories, about
which very little is known.
In this paper we construct simple models of conformal technicolor
based on QCD in the ``conformal window.''
We also propose a practical method to measure the scaling
dimension of the operator $\bar{\psi}_L \psi_R$.
This gives a new lattice probe of conformal dynamics,
and measures one of the most important parameters in
conformal technicolor models.
The reader interested only in the lattice aspects may wish to
skip to section 3.

To motivate our discussion,
we begin by reviewing the problems with strong electroweak symmetry
breaking and how they are addressed in conformal technicolor.
We begin with fermion masses.
The most straightforward approach to generate fermion masses
is to couple the standard model fermions directly to the 
operator $\Phi$ whose VEV breaks electroweak symmetry:
\beq[Oflavor]
\De\scr{L} \sim \frac{1}{M^{d-1}} \bar{Q}_L q_R \Phi,
\eeq
where $d$ is the dimension of $\Phi$.
We will be interested in the case where $\Phi$ is a technifermion
bilinear of the form $\bar\psi_L \psi_R$.
If the theory has QCD-like dynamics then $d = 3$,
but we will see that smaller values are allowed.
The term \Eq{Oflavor} is therefore a higher-dimension interaction
that requires new physics at a scale $M$.%
\footnote{%
Alternatively, it is possible that 
standard model fermions get their mass by
mixing with composite states \cite{Kaplancomp},
an idea that arises naturally in the context of
Randall-Sundrum models \cite{RS} with fermions in the
bulk \cite{RSbulkf}.
These models are ``dual'' descriptions of strongly coupled
conformal theories \cite{RSdual}.
These have a somewhat different set of flavor problems
(reviewed for strongly coupled
small-$\nc$ theories in \Ref{CTC}), and we
will focus on models based on \Eq{Oflavor}.}

The flavor problems are of two kinds.
First, there is the problem of whether 
the new physics at the scale $M$ also
generates operators of the form
\beq[qqqq]
\De\scr{L} \sim \frac{1}{M^2} (\bar{Q}_L q_R)^2.
\eeq
These potentially give rise to flavor-changing neutral currents.
This is a serious problem in ``extended technicolor''
models in which the operators \Eq{Oflavor} are generated
by massive gauge bosons coupling the technifermions and
standard model fermions \cite{ETC}.

The second flavor problem is the fact that the 
operator \Eq{Oflavor} gets strong at a scale
that is not far above the TeV scale.
For QCD-like technicolor this scale is $\sim 3\TeV$, while for
a walking theory with $d = 2$ it is $\sim 10\TeV$.
This means that in this framework the dynamics that generates
the top quark mass cannot be separated from the dynamics that
breaks electroweak symmetry.
Models in which the interactions responsible for fermion
masses are strong and play a role in electroweak symmetry
breaking are discussed in \Refs{topcolor}.
The required strong dynamics is similar to those in the
Nambu--Jona-Lasinio model \cite{NJL}, and
requires fine-tuning and additional dynamical
assumptions \cite{NJLgauge}.

\begin{figure}[t!]
\begin{center}
\includegraphics[]{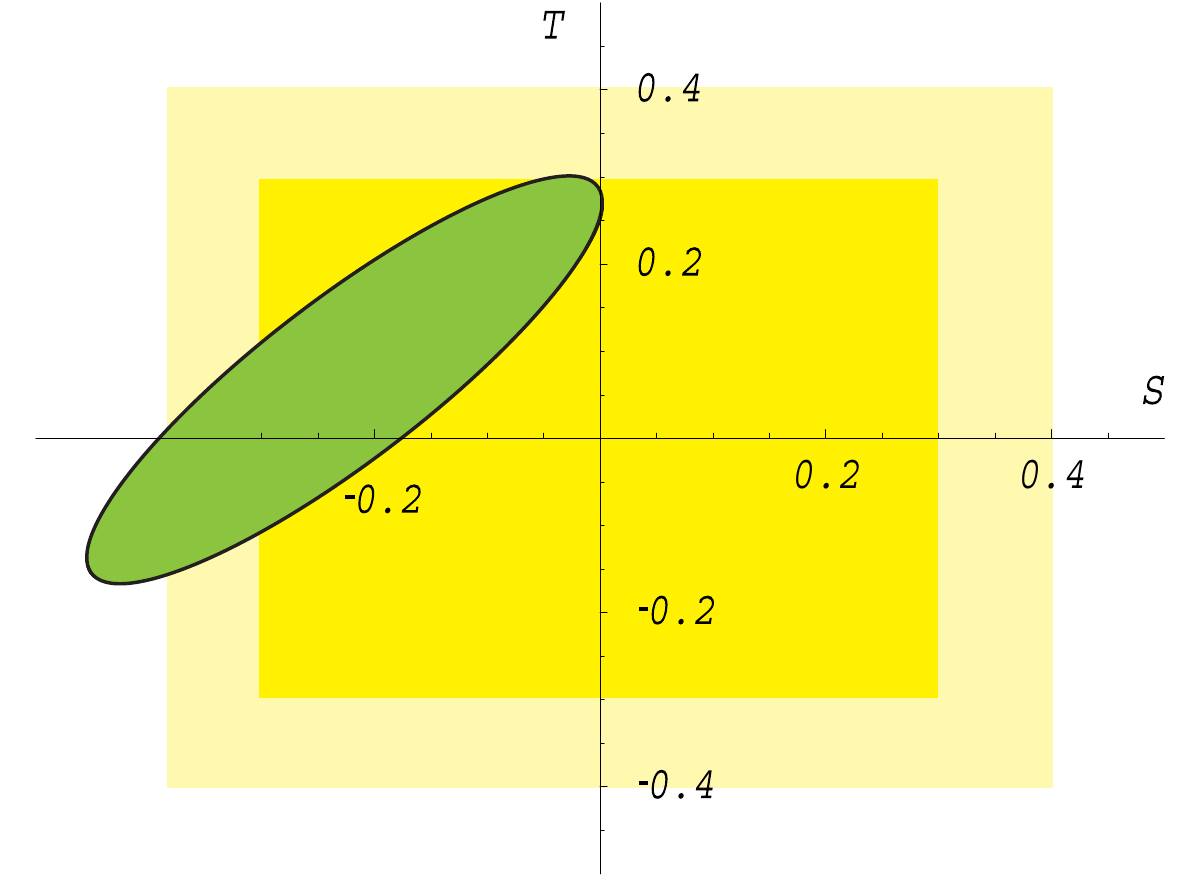}
\begin{minipage}[t]{5in}%
\caption[]{95\% confidence level
constraints on
$S$ and $T$ from precision electroweak data.
The constraints are computed
for a reference Higgs mass of $500\GeV$, so the values
correspond to the high-energy contribution in a
strongly coupled theory.
The shaded squares correspond to the 
estimates $S, T \sim \pm 1/\pi$
for a strongly coupled theory.}
\label{fig:st}
\end{minipage}
\end{center}
\end{figure}

Strong electroweak symmetry breaking
also has potential problems with precision electroweak constraints.
Using only ``\naive dimensional analysis'' \cite{NDA}
to estimate the contribution from the strongly-coupled sector,
the situation is illustrated in Fig.~\ref{fig:st}.
From this alone we would conclude that fitting the data
does not require any tuning.
However, this requires that the strongly coupled theory
give $S < 0$.
For QCD-like theories one finds quite robustly that
$S > 0$ \cite{STC}.
Extended technicolor theories can 
have negative contributions to $S$
from pseudo-Nambu-Goldstone bosons \cite{LSSTC}.
Also, studies of ``walking'' technicolor suggest that the strong-coupling
contribution to $S$ is significantly smaller than in QCD-like
technicolor, and may even be negative \cite{SWTC}.
The theories we discuss below have dynamics similar to
walking, and therefore some of these mechanisms may reduce
$S$ compared with the estimate from scaling up QCD.
At minimum, we can make the conservative statement that the
theories discussed here do not have QCD-like dynamics,
and there is nothing that excludes the possibility that they
have small $S$, or even $S < 0$.
We conclude that a good electroweak fit is possible
due to non-QCD-like strong dynamics at the TeV scale
and/or new light states below the TeV scale.

Of the three problems reviewed above,
the tension with precision electroweak physics
is generally viewed as the most problematic.
It is indeed a striking fact that a weakly coupled extension
of the standard model with a light Higgs boson
(\eg\ SUSY) naturally satisfies all
electroweak constraints.
However, the minimal supersymmetric standard model
suffers from the fact that the Higgs
is \emph{too} light, requiring either fine-tuning or
significant amount of non-minimal structure.
An alternative is ``little Higgs'' models \cite{LittleHiggs}
and their cousins \cite{twinfold} that solve only the
``little'' hierarchy problem associated with precision
electroweak data.
These also require carefully
constructed new sectors, and achieve only technical naturalness
up to scales of order $10\TeV$.
Given these difficulties, it seems worthwhile to explore the
alternative possibility
that the agreement of the standard model with precision
electroweak data is somewhat accidental.

The problem with flavor-changing neutral currents
depends on UV physics, and is very model-dependent.
One simple solution is that the operators \Eq{Oflavor}
are generated by scalar Higgs exchange in a theory
with SUSY broken at a scale $\gsim 10\TeV$, an idea
known as ``bosonic technicolor'' \cite{BosonicTC}.
Since all 4-fermion operators are proportional to Yukawa
couplings, the operators \Eq{qqqq} are diagonal in the 
same basis as \Eq{Oflavor} (up to small renormalization
group effects) and flavor-changing neutral currents are
sufficiently small.
It may appear implausible to have both SUSY and
technicolor, but it is worth remembering that weak-scale SUSY
also has a flavor problem from squark mass mixing.
In bosonic technicolor, SUSY and technicolor each solve
the flavor problem of the other.
SUSY brings flavor structure down from the far UV, while
strong dynamics is responsible for breaking electroweak
symmetry while suppressing the flavor-violating
effects of squark masses.

This leaves us with the top quark.
We view this as the most serious problem, since it
requires new physics near the TeV scale.
This is the problem that motivated conformal technicolor.
In this framework the electroweak
symmetry breaking sector is near a strongly-coupled
conformal fixed point above the TeV scale.
In a general conformal field theory, the dimension of a
scalar operator such as $\Phi$ is constrained only to be
larger than $1$ \cite{CFTdimbound}.
For example, if $d = 1.5$ the top quark operator gets
strong at a scale of order $30\TeV$.
If the flavor physics is a bosonic technicolor theory,
we only require the top quark coupling to be weakly
coupled at $10\TeV$, in which case $d \lsim 1.5$ is
probably sufficient.

Can we really have such large anomalous dimensions?
If $d < 2$ then na\"\i{}vely the singlet operator
$\Phi^\dagger \Phi$ has dimension $2d < 4$.
Then the theory has a relevant operator that cannot be
forbidden by any symmetry from appearing in the Lagrangian, 
and the fixed point is not IR stable.
The correct general formulation
is that $\Phi^\dagger \Phi$
is the singlet operator of lowest dimension in the OPE
of $\Phi$ and $\Phi^\dagger$.
This operator need not have twice the dimension of $\Phi$
due to anomalous dimensions.
In perturbation theory the anomalous dimensions are small,
but they can be order 1 in a strongly coupled theory.
We cannot get arbitrarily close to $d = 1$, because
in the limit $d \to 1$
$\Phi$ becomes a weakly coupled scalar field \cite{CFTdimbound}
and the dimension of $\Phi^\dagger \Phi$ approaches 2.
We also cannot have $d < 2$ in a large-$\nc$ theory,
because in these theories the dimension
of $\Phi^\dagger \Phi$ is equal to $2d + \scr{O}(1/\nc)$ 
even at strong coupling.
The same phenomenon occurs in Randall-Sundrum
models, which can be viewed as ``dual'' descriptions of
large-$\nc$ conformal field theories \cite{CTC}.

Studies of walking technicolor using the gap equation
\cite{gapequ} also find $d \ge 2$ \cite{walking,walkinggap}.
In the gap equation the condensate $\avg{\bar\psi \psi}$
forms precisely when the dimension of the operator
$\bar\psi \psi$ approaches 2.
The gap equation is an uncontrolled approximation
that keeps only leading effects in the large-$\nc$ limit,
and it is therefore not surprising that it reproduces
the large-$\nc$ relation between the dimensions of
$\bar\psi \psi$ and $(\bar\psi \psi)^2$.


The limit $d \to 1$ can be studied rigorously using
general properties of conformal field theories.
It was shown in \Ref{RR} that the dimension of
$\Phi^\dagger \Phi$
is $2d + \scr{O}((d - 1)^{1/2})$ as $d \to 1$ \cite{RR}.
This suggests that the anomalous dimension of $\Phi^\dagger \Phi$
may be large even if $d$ is small.
\Ref{RR} also gives a rigorous bound on the lowest dimension
operator appearing the OPE of $\Phi^\dagger \Phi$, but the
methods cannot distinguish the singlet from the non-singlet
operators, so this does not give a bound on $d$.

We conclude that at present there 
is no known reason that an IR stable
fixed point cannot have scalar operators with $d < 2$.
Such theories would significantly improve the prospects
of overcoming the most serious problem with
strong electroweak symmetry breaking, namely the top quark mass.

\section{A Simple Model}
We argued above that a conformal field theory with $d < 2$
must have both strong coupling and small $\nc$.
A natural candidate theory is therefore $SU(\nc)$ gauge
theory (with $\nc = 2$ or $3$ say)
with $\nf$ flavors in the conformal window.
It has been long known 
that $SU(\nc)$ QCD is conformal when the 1-loop beta
function is sufficiently small, at least for large $\nc$
\cite{BanksZaks}.
In supersymmetric QCD the work of Seiberg shows that
conformal fixed points are much more generic:
there is a large ``conformal window''
$\frac 32 < \nf / \nc < 3$ that exists even for small $\nc$.
This is a strong hint that conformal fixed points may be
a feature of non-supersymmetric QCD over a wide range
of $\nc$ and $\nf$.
Lattice studies have investigated this question and have
found evidence for conformal fixed points.
Iwasaki {\it et.\ al.}\ find evidence for an
IR fixed point for $\nc = 3$ and $7 \le \nf \le 16$ \cite{Iwasaki}.
Appelquist {\it et.\ al.}\ find evidence for a fixed point
at $\nf = 12$ but not at $\nf = 8$ \cite{AppFlemNeil},
and Deuzeman {\it et.\ al.}\ also find that
$\nf = 8$ is in the confining phase \cite{Deuzeman}.
The non-perturbative renormalization group study of
\Ref{nonPRG} suggests that the fixed point exists
for $12 \le \nf \le 16$.
These results are encouraging for the existence of a strong
conformal phase, although the discrepancies
clearly need to be understood.

These can be compared with an extrapolation
from the conformal window of SUSY QCD,
which suggests a conformal window
\beq
3.25 \lsim \frac{\nf}{\nc} \lsim 5.5.
\eeq
for non-supersymmetric QCD \cite{AppTernPhase}.
This leads us to expect strongly-coupled fixed points
for non-supersymmetric $SU(2)$ near $\nf = 9$,
and for $SU(3)$ near $\nf = 13$.


We therefore consider an $SU(\nc)$ gauge theory with $\nf$
flavors assumed to be a strong conformal field theory.
We weakly gauge this theory with the standard model so that
2 of the flavors form a minimal technicolor
sector \cite{minTC}
while the remaining flavors are electroweak singlets.
That is, under the gauge group
$SU(\nc)_{\rm CTC} \times SU(2)_{\rm W} \times U(1)_{\rm Y}$
the fields are
\beq\bal
\psi_L &\sim (\nc, 2)_0,
\\
\psi_R &\sim (\nc, 1)_{\frac 12},
\\
\psi'_R &\sim (\nc, 1)_{-\frac 12},
\eal\eeq
together with $\nf - 2$ copies of
\beq
\bal
\chi_L &\sim (\nc, 1)_0,
\\
\chi_R &\sim (\nc, 1)_0.
\eal
\eeq
We also assume that the theory has an explicit mass term for
the electroweak singlet $\chi$ fields
\beq
\De \scr{L} = m_\chi \bar\chi_L \chi_R + \hc
\eeq
(For $\nc = 2$ there is no invariant difference between
$\chi_L$ and $\chi_R$ except as defined by this mass term.)
Theories of this kind have been considered previously
\cite{partgauge} with different dynamical assumptions
(see section 4 below).

This theory is assumed to be at a conformal
fixed point above the TeV scale.
The explicit mass term is a relevant operator that
forces the theory to flow away from the fixed point at the 
TeV scale.
Below the TeV scale, we expect the theory to be in the same
universality class as the theory with $2$ massless flavors.
That is, we expect that the theory confines and breaks chiral symmetry
just like minimal technicolor.
A heuristic picture is that below the scale of the $\chi$
mass term the gauge coupling is no longer prevented from
blowing up by the $\chi$ fermions, and therefore confines
just like the theory with 2 massless flavors.
(This sort of straightforward decoupling is known to hold in 
supersymmetric QCD.)
Of course, this is another dynamical assumption that should be
checked.

Because the theory is in a nontrivial fixed point
the operator $\bar\chi_L \chi_R$ will have a nontrivial
scaling dimension $d \ne 3$ in general.
The theory has an anomaly-free $U(\nf)$ flavor
symmetry that means that this has the same dimension
as the operators
$\bar\psi_L \psi_R$ and $\bar\psi_L \psi'_R$
that generate the top quark mass.
We assume that $d \lsim 1.5$ to allow the top quark to be
weakly coupled to the strong sector.

This means that the $\chi$ mass term is a relevant operator.
A dimensionless measure of the strength of the coupling
$m_\chi$ at the scale $\mu$ is
\beq
\hat{m}_\chi(\mu) \sim \frac{m_\chi(\mu_0)}{\mu_0}
\left( \frac{\mu_0}{\mu} \right)^{4 - d}.
\eeq
Here $\mu_0$ is a reference renormalization scale.
The scale $\mu$ where $\hat{m}_\chi \sim 1$ determines the scale
of conformal symmetry breaking:
\beq[EWbreakscale]
\La_{\rm CTC} = \mu_0
\left( \frac{m_\chi(\mu_0)}{\mu_0} \right)^{1/(4 - d)}.
\eeq
Since the coupling is already strong at the fixed point,
we expect the transition to the new phase to be rapid.
Since we assume that this new phase confines like 2-flavor QCD,
we identify $\La_{\rm CTC} \sim \mbox{TeV}$
with the scale of strong electroweak symmetry breaking.

In this model, the scale of electroweak symmetry breaking is 
put in ``by hand'' in terms of the value of the $\chi$ mass
term.
The important point is that the smallness of the $\chi$
mass term is protected by a chiral symmetry, so this is
a perfectly natural solution of the hierarchy problem.
In fact, this is closely analogous to the mechanism
that protects the hierarchy in supersymmetric models.
There is an extension of spacetime symmetry (conformal symmetry)
that forbids masses for fields in the symmetry breaking sector,
and this symmetry is explicitly but softly broken to give
an explanation for the smallness of the weak scale compared
to \eg\ the Planck scale.

\section{Conformal Dynamics on the Lattice}
The conformal technicolor models described above rely on the
conformal dynamics of $SU(N_c)$ gauge theories with $N_f$ flavors.
The most important of these are
the existence of the conformal fixed point
for suitable values of $N_f$ and $N_c$
and the dimension $d$ of the order parameter
$\bar\psi_L \psi_R$.
We now describe how these can be addressed on the lattice.

We consider the theory with nonzero fermion masses
for all flavors.
This completely breaks both chiral and conformal symmetry,
allowing a practical test of the strong $SU(N_c)$ dynamics
we we will see.
To begin with, we take all masses to have a common
value $m$.
The scales in the lattice simulation must satisfy
\beq[latticehierarchy]
L^{-1} \ll m_{\rm had} \ll \La \ll a^{-1},
\eeq
where $L$ is the lattice size,
$m_{\rm had}$ is a low-lying hadron mass (see below),
$\La$ is the scale where the gauge coupling gets strong,
and $a$ is the lattice spacing.
The first two inequalities are required so that
the physics at the hadronic mass scale is probed in the simulation;
the final inequality is required to ensure
that the gauge coupling at the lattice scale is
weak, and therefore the lattice simulation is probing the 
desired continuum limit.
Lattice simulations have a limited
dynamical range, and these hierarchies will not be large
in practice.

To distinguish the 
confining and conformal phases of gauge theory, we use
the hadron masses.
We first discuss the more familiar confining phase.
Assuming that the quark masses are sufficiently small,
the theory confines at the scale $\La$, and spontaneously
breaks the 
$SU(N_f) \times SU(N_f)$ symmetry down to $SU(N_f)$.
This results in $N_f^2 - 1$
pseudo Nambu-Goldstone bosons.
They will have mass related to the quark mass $m$ by
\beq
m_\pi = c_\pi (\La m)^{1/2}
\eeq
for small $m$.
Other mesons such as the $\rho$ will have masses independent
of $m$ at leading order:
\beq
m_\rho = c_\rho \La.
\eeq
The constants $c_\pi$ and $c_\rho$
depend on the precise definition of
$m$ and $\La$, but the scaling with $m$
is unambiguous, since both $m$ and $\La$ are 
well-defined up to a multiplicative constant.

In the conformal case, the theory enters a conformal
phase at scales below $\La$.
The quark mass term breaks the conformal
symmetry explicitly, and we expect that
the theory below the conformal symmetry
breaking scale has a nonzero mass gap.
Note that the global symmetry of this model is a vectorlike
$U(\nf)$ symmetry, which cannot be spontaneously broken
\cite{VafaWitten}.
Therefore, there can be no massless Nambu-Goldstone
bosons in this theory.
The absence of a chiral symmetry also means that
we do not expect massless composite fermions in this
theory.%
\footnote{%
It is possible that the theory flows to a different
conformal fixed point, in which case there is no
mass gap.
This will show up as a power law volume dependence
of correlation functions (see below).}

If conformal symmetry is broken at scales smaller
than $\La$, the conformal symmetry breaking scale is
effectively the only scale in the theory.
All hadron masses in this theory are therefore
proportional to this scale, which is given by
\Eq{EWbreakscale}:
\beq[Mconformal]
\mbox{conformal:}\  
m_{\pi, \rho} = \tilde{c}_{\pi,\rho}
\La \left( \frac{m}{\La} \right)^{1/(4 - d)}.
\eeq
As in the confining phase, the scaling with $m$
holds independently of the precise
definition of $m$ and $\La$.

We see that measurement of hadron masses in the theory with
quark masses gives a direct test of the conformal dynamics.
What is important is the scaling of the hadron masses
with the quark masses, with the hadronic scale $\La$ held
fixed.
As long as the quark masses are sufficiently small,
this is equivalent to holding the bare (lattice) coupling
fixed.
The scaling with the quark masses is then well-defined
as long as the quark masses are only multiplicatively
renormalized.

One way to measure the hadron masses is to use 2-point
functions $\avg{\scr{O}^\dagger(0) \scr{O}(x)}$,
where $\scr{O}$ is an operator with the quantum numbers of
the hadron of interest.
As long as the operator $\scr{O}$ does not have vacuum quantum
numbers, the 2-point function is dominated by the lightest
particle $\varphi$ with the quantum numbers of $\scr{O}$
for $|x| \gg 1/m_\varphi$.
For $|x| \ll L$ this is a simple exponential dependence
$\avg{\scr{O}(0) \scr{O}(x)} \sim e^{-m_\varphi |x|}$.
For $|x| \sim L$ 
the dependence on $m_\varphi$ is more complicated, but is
completely determined by the geometry.

Another possibility is to use finite-volume scaling to measure
the lightest hadron mass.
This has been used extensively in studies of
conformal systems in statistical
mechanics \cite{finitesizescaling}
and in lattice gauge theory \cite{LatticeFiniteV}.
Here we focus on expectation values of 
gauge-invariant local singlet
operators $\scr{O}$ such as $\bar\psi \psi$ or 
$\tr F^{\mu\nu} F_{\mu\nu}$.
For finite $L$, the leading volume dependent
effects come from the lightest states propagating
from one side of the lattice to the other.
(We assume that the lattice has periodic boundary
conditions, so there is a discrete translation
symmetry and therefore no preferred position on
the lattice.)
In the confining case
the matrix element $\bra \pi \scr{O} \ket \pi$
is nonzero, so the
leading finite volume effects are controlled by the single pion states.
In the conformal case, the mass of the lightest
hadron is set by \Eq{Mconformal}.
In both cases, calling the lightest hadron $\varphi$,
we have for $L \gg m_\varphi^{-1}$
\beq
\avg{\scr{O}} = C_0 + C_1 e^{-m_\varphi L}
+ \mbox{independent\ of\ }L.
\eeq
Different operators will give different values 
of $C_0$ and $C_1$, but they have the same exponential
falloff with $m_\varphi$, 
so using different operators can be used as a consistency check.
A disadvantage of this way of measuring the mass
gap is that the exponential dependence must be extracted from 
on top of a larger constant. 
Furthermore, all particles contribute to the volume
dependence, so there may be several particles with
similar masses contributing to the exponential.

Measuring the dependence of the hadron masses on the quark
mass is also useful to check the other main dynamical
assumption of conformal technicolor theories,
namely the spontaneous breaking of chiral symmetry
in the theory with 2 massless flavors.
To probe this, we give 2 of the flavors a smaller quark mass
$m' \ll m$.
If the theory behaves as we have assumed above, we have
\beq
m_\pi &\propto \left[ m' m^{1/(4-d)} \right]^{1/2},
\\
m_\rho &\propto m^{1/(4-d)}.
\eeq
Here $\pi$ is a pseudo-Nambu-Goldstone boson associated
with the breaking of the $SU(2) \times SU(2)$ chiral symmetry
of the light flavors, while $\rho$ is any other hadron.
Measuring this requires a further hierarchy of lattice scales,
and will therefore be more costly to simulate.

\section{Conformal Versus ``Walking'' Dynamics}
We have not discussed the possibility of ``walking'' dynamics
because we feel that it is less likely to occur than the conformal
dynamics we are discussing.
Recall that ``walking'' is the assumption that the gauge coupling is
large over a large range of scales,
but eventually blows up in the IR, leading to confinement
and chiral symmetry breaking.

Conformal field theory gives a well-defined
physical framework that can be used to address the question
of walking dynamics.
%
%
This has already been noted by many authors in the
literature on walking technicolor \cite{walking,walkinggap}.
%
The assumption of walking is equivalent to the assumption
that the theory is near a conformal fixed point
for a range of energies, but does not flow to the fixed point
in the deep IR.
This means that the fixed point is unstable in the IR,
\ie\ there is a relevant
operator in the Wilsonian effective
action at the scale $\La$ where the theory first becomes
strongly coupled.
The fact that the walking theory nonetheless comes close to
the fixed point means that the relevant operator
has a small coefficient in the effective Lagrangian,
and/or the dimension of the relevant operator
is close to $4$.
Either scenario requires the presence of a small parameter.

In a theory with a large number of colors $\nc$ and flavors $\nf$,
the anomalous dimensions of operators are a function of
$\nf/\nc$, which becomes effectively a continuous variable.
It is therefore plausible that in such theories there is a
singlet operator whose dimension approaches 4 at a
critical value of $\nf/\nc$.
Such a theory could indeed exhibit walking behavior.
Even then, one must assume that the relevant
operator has a suffuciently small coefficient in the Wilsonian
effective action at the scale $\La$ to come close to the
fixed point.
However, a large-$\nc$ walking technicolor theory
gives values of $S$ and $T$ enhanced by a factor of $\nc$,
and is therefore not a good candidate for a theory of
electroweak symmetry breaking.
A viable (small $\nc$)
walking theory requires an additional unexplained
small parameter in order to come close to an IR
unstable conformal fixed point.

It is for this reason that we believe that conformal dynamics
is more robust than walking dynamics.
Ultimately, however it is experiment that must decide.
A practical method of looking for walking dynamics
is the Schr\"odinger functional technique 
\cite{Schrodingerfunctional} used by \Ref{AppFlemNeil}.
This  computes a non-perturbative running gauge coupling
where the scale parameter is the volume of the system,
and can be evaluated over a large range of scales.
It is worth noting that the results of \Ref{AppFlemNeil}
obtained with this method for $N_c = 3$, $N_f = 12$
shows no sign of exiting the fixed point,
and therefore indicate conformal rather than walking dynamics.

\section{Conclusions}
We have argued that strong conformal dynamics can give an elegant
solution of the hierarchy problem, while addressing the problems
of strong electroweak symmetry breaking.
Whether these ideas have anything to do with reality will be
decided by experiments, both at the LHC and on the lattice.
Investigating strong conformal dynamics on the lattice is also
an interesting problem in its own right.
We have suggested a simple method using the scaling dependence of 
hadron masses on the quark mass to find evidence for conformal dynamics
and measure the scaling dimension of the operator $\bar\psi \psi$.
We hope that these methods will prove useful and lead to further
investigation in this area.

\section*{Acknowledgements}
I would like to thank the organizers of the 
``Lattice Gauge Theory for LHC Workshop'' at
Lawrence Livermore Laboratory May 2008,
for an invitation to speak
that stimulated this work.
I thank N. Christ and K. Holland for helpful comments on lattice
simulations.
I thank S. Chang and T. Okui for comments on the manuscript.


\end{document}